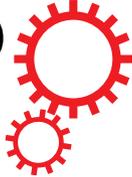

# SCIENTIFIC REPORTS

**OPEN**

# Observation of Transient Overcritical Currents in YBCO Thin Films using High-Speed Magneto-Optical Imaging and Dynamic Current Mapping

Frederick S. Wells[1], Alexey V. Pan[1,5], Igor A. Golovchanskiy[1,2,3], Sergey A. Fedoseev[1,4] & Anatoly Rozenfeld[4]

The dynamics of transient current distributions in superconducting $YBa_2Cu_3O_{7-\delta}$ thin films were investigated during and immediately following an external field ramp, using high-speed (real-time) Magneto-Optical Imaging and calculation of dynamic current profiles. A number of qualitatively unique and previously unobserved features are seen in this novel analysis of the evolution of supercurrent during penetration. As magnetic field ramps up from zero, the dynamic current profile is characterized by strong peaks, the magnitude of which exceed the conventional critical current density (as determined from static current profiles). These peaks develop close to the sample edges, initially resembling screening currents but quickly growing in intensity as the external field increases. A discontinuity in field and current behaviour is newly observed, indicating a novel transition from increasing peak current toward relaxation behaviour. After this transition, the current peaks move toward the centre of the sample while reducing in intensity as magnetic vortices penetrate inward. This motion slows exponentially with time, with the current distribution in the long-time limit reducing to the expected Kim-model profile.

The field and current profiles of $YBa_2Cu_3O_{7-\delta}$ (YBCO) and other superconducting samples are well understood for different magnetic states and histories under static conditions[1–10]. However, many of the most interesting magnetic effects in superconductors are transient and occur at extremely high speed: the spectacular dendritic flux avalanches appear to occur nearly instantaneously - their actual speed could be between 5 km/s[11] and 180 km/s[12]; while smaller localised flux jumps can occur over time-scales less than 0.1 s[13].

The final penetration depth of flux into YBCO films has been shown to be dependent on the ramping rate of applied field[14,15], suggesting that transient flux dynamics during penetration could be a strong determining factor in the measured final field state of any superconducting sample.

In order to investigate such transient and dynamic events in superconductors, new high-speed magnetic measurement techniques are required. Most local magnetic measurements on superconductors are carried out using various scanning techniques[16], which are fundamentally limited in speed due to their scanning mechanism[17]. On the other hand, laser pulsed magneto-optical imaging is able to obtain pairs of images with nanosecond-scale variable delay[12], but it cannot record continuously in order to fully investigate superconducting magneto-dynamics.

[1]Institute for Superconducting and Electronic Materials, University of Wollongong, Northfields Avenue, Wollongong, NSW 2522, Australia. [2]Laboratory of Topological Quantum Phenomena in Superconducting Systems, Moscow Institute of Physics and Technology, State University, 9 Institutskiy per., Dolgoprudny, Moscow Region, 141700, Russia. [3]Laboratory of Superconducting Metamaterials, National University of Science and Technology MISIS, 4 Leninsky prosp., Moscow, 119049, Russia. [4]Center for Medical & Radiation Physics, University of Wollongong, Northfields Avenue, Wollongong, NSW 2522, Australia. [5]National Research Nuclear University MEPhI (Moscow Engineering Physics Institute), 31 Kashirskoye Shosse, 115409, Moscow, Russian Federation. Correspondence and requests for materials should be addressed to A.V.P. (email: pan@uow.edu.au)

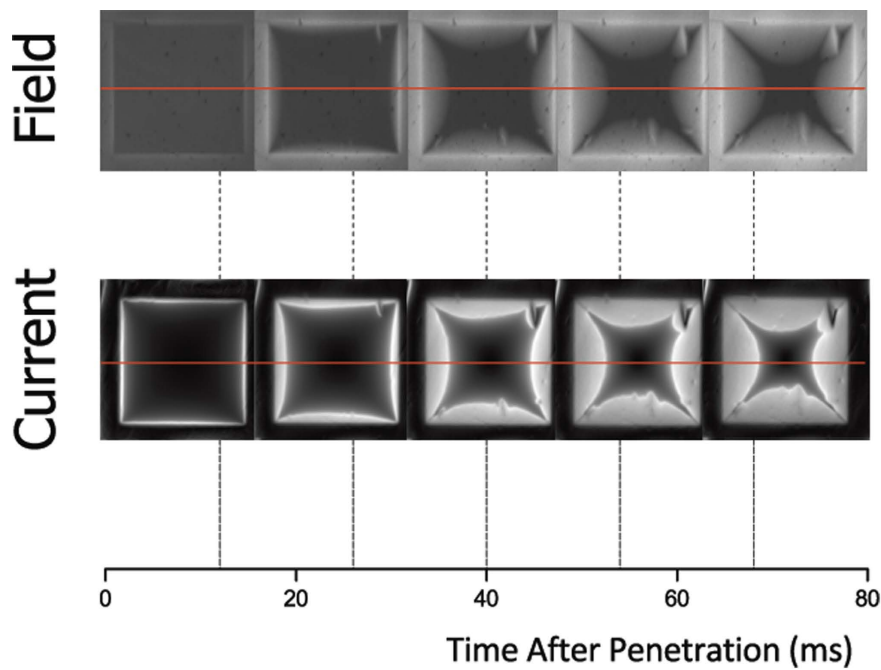

**Figure 1. Timeline of magnetic field and current in a 3 mm square of thin film YBCO during field penetration.** External field ramp begins at 0 ms and is stable at 0.1 T after approximately 30 ms, but field and current in the sample continue to evolve. Temperature is maintained at 7 K. The red line shows the central line used for one-dimensional field and current profiles.

More recently, the high-speed magneto-optical video (MOV) technique has been developed in order to continuously record changes in the magnetic state of superconducting samples at rates up to 30 kHz[18].

In this work, we investigate the behaviour of magnetic flux and current in YBCO thin films in a dynamic, time-dependent manner as the external field is varied. This is achieved by quickly ramping the applied magnetic field from one state to another and observing the resulting local changes in field in the sample using real-time MOV.

We then investigate the evolution of the current distribution as the field changes, using a Biot-Savart inversion procedure[1,19–21] on each frame of MOV sequentially, allowing millisecond-scale precision. We have shown elsewhere that the Biot-Savart law is valid under dynamic conditions for our measurements, within a close approximation[22].

We find that during initial flux penetration the current distribution does not show any plateau-like features (as would be expected for a static current distribution[1,3,8]), but instead peaks strongly near the sample edges. The magnitude of currents in this transient state appear to exceed the static critical current density, as determined by applying the Kim critical-state model to these measurements in the long-time limit[3–6]. Such a distribution can be expected under dynamic conditions, as seen in pulsed field magnetisation simulations for bulk samples during external field increase, where a similar field evolution was linked to heat propagation in the sample during rapid flux entry[23,24]. A qualitatively similar transient current distribution was also seen in simulation for AC applied fields[25].

The behaviour of field and current is radically different once the external field becomes stable. Current relaxation after a change in field/current conditions has been investigated over longer time-scales[26], with relaxation toward a time-independent current profile usually occurring in an exponential manner[27], which is confirmed in this study. However, with high-speed measurements, we observe a distinct transition from strongly-peaked increasing currents toward relaxation behaviour, which has never previously been reported.

The shape of the current profile finally settles to a distribution resembling that predicted by the Kim model[3–6], with field-dependent critical currents described by:

$$J_c(B) = J_{c0}\frac{1}{1 + B/B_0} \quad (1)$$

where $J_{c0}$ is the zero-field critical current and $B_0$ is a magnetic field constant.

## Methods

YBa$_2$Cu$_3$O$_{7-\delta}$ films were produced by pulsed laser deposition described in refs 28, 29, and given the square shape by ion beam etching using the procedure described in ref. 22. These films have measured critical current density of $3.5 \times 10^{10}$ A/m$^2$ at 77 K[28,29].

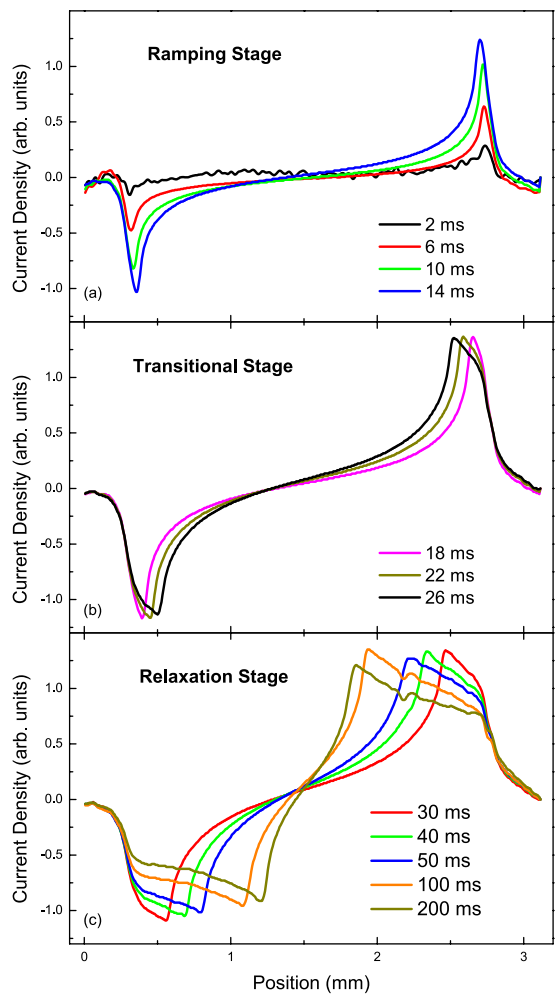

**Figure 2.** (**a**) Evolution of the current profile on the central line of the film during the ramping stage, as field ramps from zero to 0.01 T over $(27 \pm 3)$ ms. (**b**) Current evolution in the brief intermediate stage, when external field has not reached its maximum value but peak position begins to shift as in the relaxation stage. (**c**) The subsequent current evolution in the relaxation stage, which occurs over the following 200 ms as the sample approaches equilibrium. The small dip around 2.1 mm is due to the influence of a sample defect. Vertical lines show the expected position of the sample edges.

After cooling to 7 K at zero field, magnetic field was applied to the samples with a maximum ramping speed of $(\dot{B}_a = 8.1 \pm 2.3)$ T/s up to a maximum of 0.1 T[22]. The resulting flux dynamics in this penetration event were investigated at 500 frames per second (fps), with a system capability of up to 120000 fps, using the high-speed camera and MOV apparatus described in ref. 22. Cernox SD thermometers were mounted at the tip of the cold finger (next to the sample) and at its base (on the heat exchanger). These showed no noticeable temperature change on magnetic field ramping, which may have occurred due to eddy currents.

The magneto-optical indicator films used consist of a thin single-crystalline layer of Faraday-active bismuth-substituted yttrium iron garnet on a transparent gadolinium gallium garnet substrate[30], whose production and properties has been described in refs 31, 32]. A 125 nm thick Al reflective layer is deposited on top of the active layer, followed by a 120 nm capping layer which is placed face-down onto the sample leaving ~1 $\mu$m gap. Similar indicator films have been shown to have a very fast linear response to changing magnetic fields, so the indicator does not limit measurement speed in this case[22,33]. Eddy currents will be induced in the Al layer during ramping, but since this layer is thin and at a sufficiently large distance, these should not have any significant effect on the sample[34,35].

A map of the current density at each time-step was calculated using an inverse Biot-Savart procedure on each frame of the video[1,19,20]. One-dimensional profiles were extracted from the calculated current data, along the central line shown in Fig. 1 for the current timeline. This central line was chosen in order to avoid inaccuracies in calculation due to transient charges which arise close to the discontinuity lines during field ramping[22,36,37].

### Results and Discussion

Magneto-optical video showing the dynamic behaviour of field and current in the films was captured under various non-constant field conditions. Figure 1 shows field and current evolution in zero-field-cooled samples during and after a 0.1 T field is ramped up. This figure provides a selection of frames from a typical MOV, each showing a snapshot of the field distribution around the YBCO sample at a given point in time. Beneath these are snapshots of the corresponding supercurrent in the film as calculated from the instantaneous field distribution at each time step. The current images are computer-generated with brighter regions showing higher current magnitude. The brightness of each of these images is scaled relative to the maximum current for that time step.

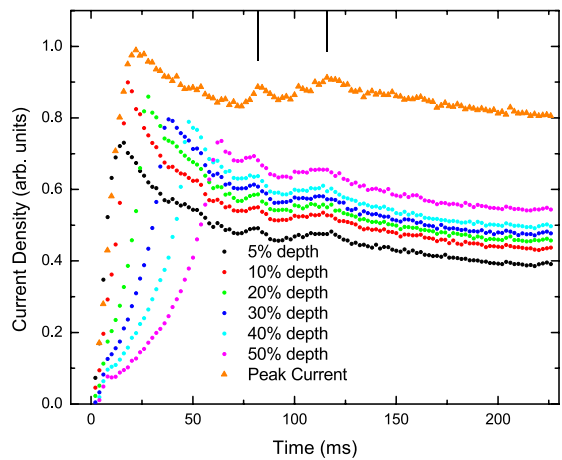

**Figure 3. Time-evolution of current density at several fixed points along the central line in Fig. 1, and of the peak value of current density.** Each curve shows a sharp increase followed by relaxation toward a final stable value. The position of each point is indicated by "depth", with 0 depth being the edge of the film, and 100% depth being the centre. Unexpected maxima that show brief increases in current during relaxation are indicated by vertical lines.

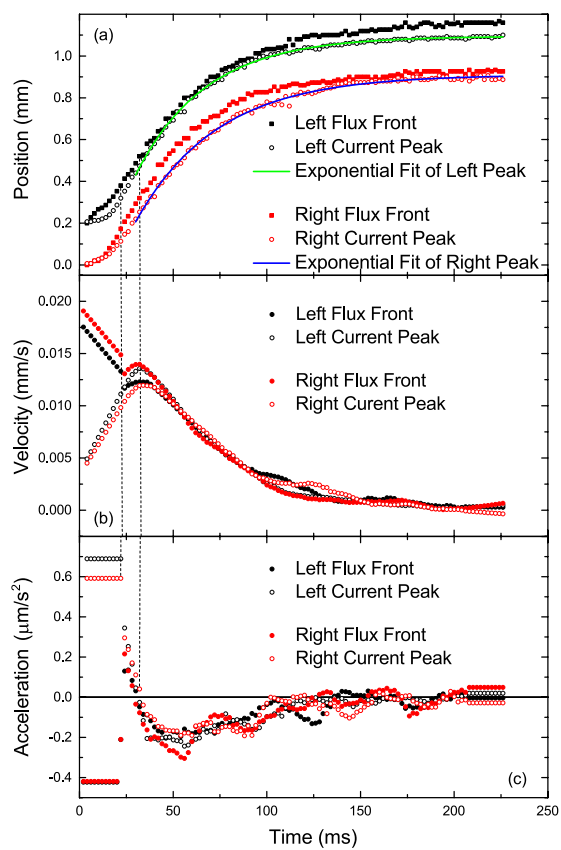

**Figure 4. Motion of the flux fronts and current peaks originating from the left and right edges of the sample.** (**a**) Position measured relative to the closest sample edge, (**b**) Velocity, (**c**) Acceleration. The left front and peak are offset by 0.2 mm on the position scale for clarity.

Each individual video provides an enormous amount of data about the current in the film. In order to analyse and quantify this information, one-dimensional slices of the calculated current distribution were taken for each time step, and these are plotted in Fig. 2.

The edges of the sample cannot be directly observed in a magneto-optical imaging experiment since the physical sample is obscured from view under the microscope by the indicator film. However, since the first non-zero currents arise due to screening of the very small initial field, we take the edges to be within a distance $\lambda$ from the outermost points at which current is first seen above the noise level. Here $\lambda$ is the effective penetration depth for YBCO thin films, which is ~180 nm for our samples[38,39].

The time-evolution of current density at several fixed points along the central line is plotted in Fig. 3, showing more clearly the numerical change in current over time. This figure also plots the maximum (peak) value of current along the central line over time. The motion of the flux fronts originating from either edge of the sample is plotted in Fig. 4 along with the peaks in current density occurring near each of these fronts.

The evolution of the current profile across the film is seen to occur in two distinct stages, which we label "ramping" and "relaxation", with an intermediate "transitional" stage in between. The following subsections describe each stage in more detail.

### Ramping Stage.
The first stage occurs during field ramping, which happens over the first 14 ms as the external field is increasing steadily. In this stage, most of current in the sample flows close to its edges. The evolution of the current profile in the ramping stage is shown in Fig. 2a.

Since the external field ramps continuously from zero, there is a finite time in which the film is in the Meissner state below its first critical field. Thus, the first currents to flow in the film are Meissner screening currents at the edges of the film within a width of the order of the effective penetration depth.

Within the first few milliseconds, these screening currents increase in magnitude as the external field increases, while the peak in current distribution shifts slightly inward. This shows that currents are no longer confined to the penetration depth and vortex entry has begun.

After 6 ms, as the applied field continues to increase, a long tail is seen in the current distribution from this peak to the sample centre. This tail in current represents a screening of the horizontal component of the field which has curved around the sample due to the high demagnetising factor of our YBCO films. At this time Meissner currents flow throughout the sample, except in the peak region, where larger currents occur.

The ramping stage is seen in Fig. 3 as the initial period for which the magnitude of the peak current is increasing.

Figure 4 reveals that the current peaks move further into the sample with constant acceleration during the ramping stage, while the flux fronts also move inward but with constant deceleration. ref. 40 gives a theoretical estimation of the vortex velocity during field ramping, which is derived on the basis of the Bean model. Following this derivation, the inward speed of vortices at a position $x$ is given by:

$$v(x) = \frac{2\pi \dot{B}_a}{\mu_0 J_c} \frac{\sqrt{x^2 - f^2}}{ln|(1-u)/(1+u)|} \quad (2)$$

where $f$ is the flux front position, $a$ is the sample half-width and $u = \frac{x\sqrt{a^2-f^2}}{a\sqrt{x^2-f^2}}$. This equation is valid throughout the penetrated region ($f \leq x \leq a$, measured relative to the sample centre), taking $x$ positive for simplicity. Applying this to our samples for maximum ramp rate and $f \geq 0.9a$ (up to 10% depth) and taking the $J_c$ value of the film at 10 K as $10^{12}$ A/m$^2$ (see the typical values measured for a similar film in Fig. 3a of ref. 22) gives vortex velocities near the flux front in the range of ~$10^{-2}$ mm/s. Strikingly, this is a similar order of magnitude to the experimentally observed flux front velocity plotted in Fig. 4c.

### Relaxation Stage.
The relaxation stage is defined as the stage when the external field has reached its stable value, which occurs after about 30 ms for our measurements[22]. The peak in supercurrent distribution moves away from the edge of the sample, as shown in Fig. 2c. In this stage, there is no longer any change in external conditions but the magnitude and position of the supercurrent peak still vary in an exponentially slowing manner, as explained below.

During the relaxation stage, the position $p$ of the current peak is seen to move toward the centre of the sample, with a sloped plateau between this peak and a shoulder, whose onset remains close to the initial peak position at the sample edge.

The plateau can be clearly distinguished after 30 ms and has a very steep gradient toward the sample edge. This gradient decreases fairly linearly with time until about 80 ms, after which the shape of the plateau region is more stable, but the current density throughout continues to decrease with time. Analysis of the plateau was carried out over the left side of the central line, since the right side is influenced by a defect in the top right corner.

The inward speed of the current peak decreases exponentially until it settles at a final position, as shown in Fig. 4a. An exponential decay curve was fitted to the motion of each current peak starting from the time the applied field was stable (30 ms). The fitting equation used was given by:

$$p = Ae^{-t/\tau} + p_0 \quad (3)$$

where $t$ is the time after field switched on, and $p_0$, $\tau$ and $A$ are the fit parameters. Both fitting curves in Fig. 4a have a coefficient of determination $R^2 > 0.996$, with $R^2 = 1$ describing the perfect fit. The time constant $\tau$ was found to be $(37.5 \pm 0.4)$ ms for the left peak and $(43.7 \pm 0.6)$ ms for the right peak (originating on opposite edges of the film). This parameter describes how quickly the current distribution relaxes to equilibrium, and is probably strongly linked to pinning strength. The physical origin of this time dependence is not fully understood and will be investigated in later works.

The motion of the current peak near each edge is plotted in Fig. 4, along with flux front motion for comparison. The flux front also displays exponentially slowing inward motion, while the current peak remains a small but non-fixed distance behind the front at all times.

The evolution of the supercurrent distribution in this stage is due to the dynamic effect of vortices moving inward in response to the large field gradient, though their motion is damped due to pinning.

During the relaxation stage, the current density at each point relaxes over time toward a stable value, as seen in Fig. 3. This value is taken to be the static critical value for each point, although the relaxation is convoluted with the change in field over time. The initial period of current increase is longer for points closer to the centre due to the constant inward motion of the flux front. The largest magnitude reduction in current density is at the 10% depth, where the maximum transient current is almost twice the asymptotic stable value. This can be attributed to a local field increase at this point during relaxation.

The current relaxation behaviour is made clearer by considering the magnitude of the current peak, since the field near the flux front is always small and fairly constant. This magnitude has a decreasing trend with time, asymptotically approaching a value which is taken to be close to $J_{c0}$.

This relaxation was expected to be monotonic, but two "bumps" are unexpectedly observed in the graph. These bumps are reproduced at every position measured, with the current density noticeably increasing universally between 72 and 82 ms and between 90 and 116 ms. These bumps may be due to oscillations in induction due to the flux wave phenomenon, which has been observed in the transient dynamic regime after switching on a magnetic field smaller than the order-disorder transition field[41] (which should be much higher than 0.1 T for YBCO samples[42]). The period of these oscillations is shorter than those seen in ref. 41, which is expected since the measurement temperature is lower.

**Transitional Stage.** One may clearly see that the evolution of current in the sample is vastly different during the relaxation stage as compared to the ramping stage. However, the current evolution begins to show features of the relaxation stage before the external field has reached a stable value. Hence we define a brief transitional stage during which the evolution of current in the sample shows features of both the ramping and relaxation stages.

During this transitional stage, the current peak begins moving away from the sample edge and its magnitude no longer increases, as seen in Fig. 2b. The shoulder near the sample edge also forms over these few milliseconds.

Over this time interval, the peak current stops its initial increase, with a stationary point seen in Fig. 3. The peak current reaches a maximum around 22 ms after penetration, which is approximately 20% higher than the asymptotic value to which the peak current converges in the long-time limit.

This transitional behaviour may be due to an intrinsic transition in the film occurring when the transient overcritical currents are no longer sustained by increasing field and therefore slowly dissipate due to vortex motion. At this point, the sample response to field ramping shifts from increasing screening-like currents which peak at the film edge toward an inward shifting of the peak current position. However, this internal factor may also be combined with the instrumental factor that the external field does not simply ramp linearly to its maximum value, but instead slows down after ~16 ms to avoid overshooting the set value[22]. The slowing of the field ramp may lengthen the transitional stage.

Evidence for a transition is also found in Fig. 4: The constant inward acceleration of both current peaks decreases discontinuously at 22 ms, as does the constant deceleration of the flux fronts. This discontinuity is also seen in the flux front velocities. After this point, the velocity of the flux front is seen to increase to a maximum at 32 ms, although it decreases monotonically outside of this 10 ms period. The velocity of the current peak also has a maximum at 32 ms, after which it decreases in the manner described by our exponential fit.

These two time-points of abruptly changing flux front and current peak behaviour are marked as dashed lines in Fig. 4, and may be taken as indications of the beginning and end of the transitional stage. These times match reasonably well with the timing of the transitional behaviour seen in Fig. 2b, and the stabilisation of external field.

## Conclusion

We have presented a novel investigation of the dynamic behaviour of field and current in YBCO thin films under a time-varying external field, using high-speed MOI. As a result, we provide in-depth analysis of the evolution of current density during flux penetration.

The experimental data shows a large overcritical peak in the dynamic current density profile very close to the sample edge, which appears within 2 ms of application of field. This deviates strongly from the current plateaus commonly seen in films under static fields. This peak increases in magnitude during the increase of external field, which we label the ramping stage. It is worth noting that previous magneto-optical measurements of YBCO strips[20] have also shown strong current peaks near the sample edges, though these were seen in static distributions and attributed to in-plane fields. This differs significantly from our measurements in which the current peaks appear only for a short time, evolving smoothly into the expected static (conventionally critical) current distribution.

We have identified a novel transition in penetration behaviour, which is seen as the external field stabilises to its maximum value. The transition is evidenced by discontinuities in peak current position and flux front velocity. At this time, the current peak stops increasing in magnitude and begins to move toward the sample centre.

After this transition, the current distribution relaxes toward equilibrium; the current peak approaches a final position closer to the sample centre, slowing exponentially and gradually reducing in intensity toward the static critical current level. A plateau of current forms during this time. It is initially very steep, but the gradient reduces over time.

The final current distribution qualitatively resembles a Kim profile[4], with current density increasing from the sample centre to a peak behind the flux front, and a plateau with downward gradient from this peak to a shoulder near the sample edge, where it drops to zero.

Therefore the accepted picture of the one-dimensional current distribution predicted by the Kim model can be seen as a limiting case which holds for static current distributions with a time-stable external field. In addition, the previously reported relaxation of current toward this stable state[26,27] occurs only after a novel transition in dynamic behaviour.

## References

1. Jooss, C., Warthmann, R., Forkl, A. & Kronmüller, H. High-resolution magneto-optical imaging of critical currents in $YBa_2Cu_3O_{7-\delta}$ thin films. *Phys. C* **299,** 215 (1998).
2. Jooss, C., Albrecht, J., Kuhn, H., Leonhardt, S. & Kronmüller, H. Magneto-optical studies of current distributions in high-$T_c$ superconductors. *Rep. Prog. Phys.* **65,** 651 (2002).
3. McDonald, J. & Clem, J. R. Theory of flux penetration into thin films with field-dependent critical current. *Phys. Rev. B* **53,** 8643 (1996).
4. Shantsev, D. V., Galperin, Y. M. & Johansen, T. H. Thin superconducting disk with B-dependent Jc: Flux and current distributions. *Phys. Rev. B* **60,** 13112 (1999).
5. Kim, Y. B., Hempstead, C. F. & Strand, A. R. Critical Persistent Currents in Hard Superconductors. *Phys. Rev. Lett.* **9,** 306 (1962).
6. Kim, Y. B., Hempstead, C. F. & Strand, A. R. Magnetization and Critical Supercurrents. *Phys. Rev.* **129,** 528 (1963).
7. Brandt, E. H. & Indenbom, M. Type-II-superconductor strip with current in a perpendicular magnetic field. *Phys. Rev. B* **48,** 12893 (1993).
8. Zeldov, E., Clem, J. R., McElkesh, M. & Darwin, M. Magnetization and transport currents in thin superconducting films. *Phys. Rev. B* **49,** 9802 (1994).
9. Polyanskii, A. A. *et al.* Magneto-optical study of flux penetration and critical current densities in [001] tilt $YBa_2Cu_3O_{7-\delta}$ thin-film bicrystals *Phys. Rev. B* **53,** 8687 (1996).
10. Polyanskii, A. A., Feldmann, D. M. & Larbalestier, D. C. Magneto-Optical Characterization Techniques. In *The Handbook on Superconducting Materials* (ed. Cardwell, D. & Ginley, D.) 1551–1567 (2003).
11. Mikheenko, P. *et al.* Nanosecond voltage pulses from dendritic flux avalanches in superconducting NbN films. *Appl. Phys. Lett.* **102,** 022601 (2013).
12. Bolz, U., Biehler, B., Schmidt, D., Runge, B.-U. & Leiderer, P. Dynamics of the dendritic flux instability in $YBa_2Cu_3O_{7-\delta}$ films. *Europhys. Lett.* **64,** 517 (2003).
13. Bobyl, A. V. *et al.* Mesoscopic flux jumps in $MgB_2$ films visualized by magneto-optical imaging. *Phys. C* **408–410,** 508–509 (2004).
14. Koblischka, M. R. *et al.* Magneto-optical observation of dynamic relaxation in $YBa_2Cu_3O_{7-\delta}$ thin films. *J. Phys. Cond. Mat.* **9,** 10909 (1997).
15. Baziljevich, M., Baruch-El, E., Johansen, T. H. & Yeshurun, Y. Dendritic instability in $YBa_2Cu_3O_{7-\delta}$ films triggered by transient magnetic fields. *Appl. Phys. Lett.* **105,** 012602 (2014).
16. Bending, S. J. Local magnetic probes of superconductors. *Adv. Phys.* **48,** 449–535 (1999).
17. Wells, F. S., Pan, A. V., Wang, X. R., Fedoseev, S. A. & Hilgenkamp, H. Analysis of low-field isotropic vortex glass containing vortex groups in $YBa_2Cu_3O_{7-\delta}$ thin films visualized by scanning SQUID microscopy. *Sci. Rep.* **5,** 8677 (2015).
18. Baziljevich, M. *et al.* Magneto-optical system for high speed real time imaging. *Rev. Sci. Instrum.* **83,** 083707 (2012).
19. Wells, F. S. *Magneto-optical imaging and current profiling on superconductors* (University of Wollongong Thesis Collection, 2011).
20. Johansen, T. *et al.* Direct observation of the current distribution in thin superconducting strips using magneto-optic imaging. *Phys. Rev. B* **54,** 16264 (1996).
21. Wijngaarden, R. J., Spoelder, H. J. W., Surdeanu, R. & Griessen R. Determination of two-dimensional current patterns in flat superconductors from magneto-optical measurements: An efficient inversion scheme. *Phys. Rev. B* **54,** 6742 (1996).
22. Wells, F. S. *et al.* Dynamic Magneto-Optical Imaging of Superconducting Thin Films. *Supercond. Sci. Technol.* **29,** 035014 (2016).
23. Fujishiro, H. & Naito, T. Simulation of temperature and magnetic field distribution in superconducting bulk during pulsed field magnetization. *Supercond. Sci. Technol.* **23,** 105021 (See Figure 4a) (2010).
24. Itoha, Y., Yanagi, Y. & Mizutani, U. Flux motion during pulsed field magnetization in YBaCuO superconducting bulk magnet. *J. Appl. Phys.* **82,** 5600 (1997).
25. Grilli, F., Lucarelli, A., Lüpke, G., Haugan, T. & Barnes, P. Dynamic Study of Field and Current Distribution in Multifilamentary YBCO Thin Films. *Proceedings of the COMSOL Conference, Boston* (See Figure 5) (2008).
26. Bobyl, A. V. *et al.* Relaxation of transport current distribution in a YBaCuO strip studied by magneto-optical imaging. *Supercond. Sci. Technol.* **15,** 82–89 (2002).
27. Warthmann, R., Albrecht, J., Kronmüller, H. & Jooss, C. Spectral distribution of activation energies in $YBa_2Cu_3O_{7-\delta}$ thin films. *Phys. Rev. B* **66,** 15226 (2000).
28. Golovchanskiy, I. A., Pan, A. V., Fedoseev, S. A. & Higgins, M. Significant tunability of thin film functionalities enabled by manipulating magnetic and structural nano-domains. *App. Surf. Sci.* **311,** 549–557 (2014).
29. Pan, A. V. *et al.* Multilayering and Ag-Doping for Properties and Performance Enhancement in $YBa_2Cu_3O_7$ Films. *IEEE Trans. Appl. Supercond.* **17,** 3585 (2007).
30. Roussel, M. Magnetic flux penetration in $MgB_2$ thin films produced by pulsed laser deposition. *Supercond. Sci. Technol.* **18,** 1391–1395 (2005).
31. Helseth, L. E. *et al.* Faraday rotation and sensitivity of (100) bismuth-substituted ferrite garnet films. *Phys. Rev. B* **66,** 064405 (2002).
32. Helseth, L. E., Hansen, R. W., Ilyashenko, E. I., Bazilijevich, M. & Johansen, T. H. Faraday rotation spectra of bismuth-substituted ferrite garnet films with in-plane magnetization. *Phys. Rev. B* **64,** 174406 (2001).
33. Wolfe, R. *et al.* High frequency magnetic field sensors based on the Faraday effect in garnet thick films. *Appl. Phys. Lett.* **60,** 2048 (1992).
34. Colauto, F. *et al.* Suppression of flux avalanches in superconducting films by electromagnetic braking. *Appl. Phys. Lett.* **96,** 092512 (2010).
35. Choi, E. M. *et al.* Dendritic magnetic avalanches in carbon-free $MgB_2$ thin films with and without a deposited Au layer. *Appl. Phys. Lett.* **87,** 152501 (2005).
36. Brandt, E. H. Electric field in superconductors with rectangular cross-section. *Phys. Rev. B* **52,** 15442 (1995).
37. Romero-Salazar, C., Jooss, C. & Hernandez-Flores, O. A. Reconstruction of the electric field in type-II superconducting thin films in perpendicular geometry. *Phys. Rev. B* **81,** 144506 (2010).
38. Pan, A. V. & Dou, S. X. Comparison of small-field behavior in MgB2, Low- and high-temperature superconductors. *Phys. Rev. B* **73,** 052506 (2006).

39. Pan, A. V. *et al.* Thermally activated depinning of individual vortices in YBa$_2$Cu$_3$O$_7$ superconducting films. *Phys. C* **407,** 10–16 (2004).
40. Schuster, T., Kuhn, H., Brandt, E. H. & Klaumünzer, S. Flux penetration into flat rectangular superconductors with anisotropic critical current. *Phys. Rev. B* **56,** 3413 (See equation 18) (1997).
41. Kalisky, B. *et al.* Spatiotemporal Vortex Matter Oscillations in Bi$_2$Sr$_2$CaCu$_2$O$_{8+\delta}$ Crystals. *Phys. Rev. Lett.* **98,** 017001 (2007).
42. Krabbes, G. *et al. High Temperature Superconductor Bulk Materials* Ch. **4,** 75–82 (Wiley, 2006).

## Acknowledgements

This work was supported by the Australian Research Council, the Australian Institute for Innovative Materials (AIIM) and the Faculty of Engineering and Information Sciences (EIS) at the University of Wollongong. The authors have also benefited from fruitful discussions with T.H. Johansen and I. Rudnev.

## Author Contributions

A.V.P. initiated and drove the work, as well as facilitated the work via the research funding. F.S.W. and A.V.P. wrote the manuscript. F.S.W. acquired and analysed the magneto-optical video, and prepared the figures. A.V.P. set up the magneto-optical apparatus and provided YBCO films. A.V.P., F.S.W. and I.A.G. discussed the results and analysed the data. S.A.F. and A.R. discussed the results. S.A.F. produced the YBCO films. A.R. contributed by providing research funding.

## Additional Information

**Competing financial interests:** The authors declare no competing financial interests.

**How to cite this article**: Wells, F. S. *et al.* Observation of Transient Overcritical Currents in YBCO Thin Films using High-Speed Magneto-Optical Imaging and Dynamic Current Mapping. *Sci. Rep.* **7,** 40235; doi: 10.1038/srep40235 (2017).

**Publisher's note:** Springer Nature remains neutral with regard to jurisdictional claims in published maps and institutional affiliations.